# Deploying AI Frameworks on Secure HPC Systems with Containers

David Brayford
*LRZ Supercomputing Centre*
Leibniz-Rechenzentrum der Bayerischen Akademie der Wissenschaften
Munich, Germany
brayford@lrz.de

Sofia Vallecorsa
CERN
*Geneva, Switzerland*
sofia.vallercorsa@cern.ch

Atanas Atanasov
*Intel Deutschland GmbH*
Feldkirchen, Germany
atanas.atanasov@intel.com

Fabio Baruffa
*Intel Deutschland GmbH*
Feldkirchen, Germany
fabio.baruffa@intel.com

Walter Riviera
*Intel Corp. UK*
Swindon, UK
walter.riviera@intel.com

*Abstract*—The increasing interest in the usage of Artificial Intelligence techniques (AI) from the research community and industry to tackle "real world" problems, requires High Performance Computing (HPC) resources to efficiently compute and scale complex algorithms across thousands of nodes. Unfortunately, typical data scientists are not familiar with the unique requirements and characteristics of HPC environments. They usually develop their applications with high-level scripting languages or frameworks such as TensorFlow and the installation process often requires connection to external systems to download open source software during the build. HPC environments, on the other hand, are often based on closed source applications that incorporate parallel and distributed computing API's such as MPI and OpenMP, while users have restricted administrator privileges, and face security restrictions such as not allowing access to external systems. In this paper we discuss the issues associated with the deployment of AI frameworks in a secure HPC environment and how we successfully deploy AI frameworks on SuperMUC-NG with Charliecloud.

I. INTRODUCTION

The convergence of high performance computing (HPC) and Artificial Intelligence (AI) is becoming increasingly important in analyzing the massive amounts of data generated by scientific experiments and numerical simulations in fields such as high energy physics, astrophysics and weather forecasting. Medical procedures for detecting diseases and personalized medicine are some of the examples for very attractive applications of that field, which can improve the daily life of all of us. Security is also one of the applications, where it is possible to detect credit card fraud and other criminal activities.

The term AI, machine learning and deep learning are used interchangeably, but they are not equivalent. AI is any intelligence that allow a machine to take independent decisions as reaction to new input data, by leveraging historical. Machine learning is a subset of AI where collections of algorithms are used to parse data, learn from it and then make a prediction. An example of machine learning is predicting house prices based on data points such as location, size, number of rooms etc. Then the algorithm creates a function out of the input parameters and adapts the unknown coefficients when it encounters more data. Finally, deep learning is a subset of machine learning and is inspired by the connections of neurons in the brain to create artificial neural networks. Unlike the brain where any neuron can connect to each other within a certain physical distance, the artificial neural networks have a finite number of layers, connections and directions of propagation.

Deep learning is been around for many years in the State-of-the-art. The "perceptron" algorithm [1] could be identified as the precursor of the current Neural Network. Although the feed-forward theory was already available in 1973, the lack of data and power to compute the mathematical functions required to instruct those type of algorithms, became an obstacle for the adoption of these tools at deployment level.

Nowadays, technology has made great progress up to the point where even generic purposes CPUs can host Deep Learning workloads. However, the size of the artificial neural networks, are getting bigger to help solve more complex "real world" problems, and the amount of data required to train it properly, are representing the new challenge for the data science community.

Among the most cost-effective solutions for pre-existing HPC infrastructures, there is the possibility of scaling out to more compute nodes and parallelize the workloads in order to speed-up the whole process.

Therefore, scalability of the computational resources is the key to solve problems, which involves complex artificial neural networks and to take advantage of the massive amount of computational power available on large scale HPC systems, such as those in the Top500 list.

Enabling data scientists to efficiently use such systems is challenging, due to the scarcity of information and software, which can accommodate both HPC and AI needs. For example, in fields such as high energy physics, scientists have started to use deep learning techniques to analyze the massive amounts of data generated during their experiments. This type of analysis cannot be done on traditional workstations and requires HPC resources and software that are able to cope with the data storage, data transfer and computation on large scale.

The purpose of this paper is to provide a mechanism to easily transition the typical AI workloads from the laptop to the supercomputer, without compromising security on the HPC systems and ensuring that the AI software packages take full advantage of the hardware and optimized libraries available on the system.

## II. RELATED WORK

In this section we introduce some of the technologies, which are commonly used to deploy AI frameworks on computer systems.

### A. Python

Python [2] is an interpreted high level programming language, which is used extensively in the fields of pre- and post-processing of data, artificial intelligence and specifically machine learning. The language is widely considered accessible for prototyping algorithms, due to the inherent constructs and the availability of optimized scientific and AI/ML specific libraries, compared to traditional programming languages such as C++. In addition, because Python is an interpreted programming language it doesn't need to be recompiled for different CPU architectures unlike C++.

The tradeoff of having flexibility, portability and ease of use is some loss in performance over code developed in C++. However, this is somewhat mitigated by using vendor optimized scientific, numeric libraries and frameworks such as NumPy, SciPy, Tensorflow and others. The Intel® distribution for Python* [3] is a free downloadable package which enables scientists to take advantage of the productivity of Python, while taking advantage of the ever-increasing performance of modern hardware. Furthermore it provides optimized implementations of NumPy and SciPy which leverage the Intel® Math Kernel Library to achieve highly efficient multi-threading, vectorization, and memory management.

However, the way Python is employed in machine learning it often requires a connection to the internet to install packages. So deploying AI software on a secure HPC system without a connection to the internet is problematic, as it requires a local repository to be set up, which is accessible from the secure system. In addition, a single instance of Python is not suitable for multi-user, multi-framework usage due to the automatic upgrading and downgrading of dependency packages. For example, installing TensorFlow followed by Caffe on the same Python instance via pip install will result in TensorFlow to no longer work, because Caffe shares some of the dependencies with TensorFlow, but requires different versions. Therefore, the automated installation mechanism will upgrade and downgrade some or all of the shared dependencies to the version required by the newly installed package and potentially break the previously installed packages.

### B. User Defined Software Stack UDSS

With the increasing demand on more flexible execution models from AI researchers the standard methods to offer software on HPC systems (i.e. predefined modules with specific software packages) becomes unpractical for the more dynamic AI software stack. User Defined Software Stacks as defined by [7] combined with the recently introduced user namespaces in Linux offer a solution how those can be realized in the form of containers without sacrificing security of the HPC cluster.

The UDSS concept helps to deal with problems such as handling dependencies, frequency software updates, support of other Linux environments (i.e. many AI applications are implemented under Ubuntu Environment).

User namespaces in Linux (since Linux Kernel 3.8) allow the execution of privileged operations without escalating permissions up to root. All children processes do not have any control over the parent processes. Mappings for UIDs and GIDs between parent and children processes make sure that operations are executing in a safe way by using the original ID of the user.

### C. Docker

Docker [4] is considered an industry standard container that provides the ability to package and run an application in an isolated environment.
However, Docker was not designed for use in a multi-user HPC environment and has significant security issues, which enables the user inside the Docker container to have privileged (root) access on the host systems network filesystem, making it unsuitable for HPC system. Also, Docker uses cgroups to isolate containers, which conflicts with Slurm scheduler, which also uses cgroups to allocate resources to jobs and enforce limits.

### D. Singularity

Singularity [5] was developed at LBL to be a containerization solution for HPC systems and supports several HPC components such as resource managers, job schedulers and contains built in MPI features. Docker images can be imported into Singularity without having Docker installed or requiring privileged access. Although Singularity has been developed to run in a non-privileged namespace, security issues have arisen on a test system at LRZ where users have escalated their privileges and the system had to be taken out of service. An example of security issues includes allowing unprivileged users to request that the kernel interpret arbitrary data as a file system and it is possible for Singularity to run directly on "bare metal" and stop, start and restart services, which could possibly bring down the file system. As a result of these security breaches and concerns, LRZ no longer allows Singularity on any of the systems.

### E. Shifter

Shifter [6] developed at NERSC in collaboration with Cray to enable Docker images to be securely executed on a HPC ecosystem. Shifter works by converting Docker images to a common format that can then be distributed and launched on HPC systems. Shifter works by enabling users to convert the Docker images to a flattened format that are directly mounted on the compute nodes using a loopback device.

Shifter appears to be a good choice for conventional HPC batch queued infrastructure. That also provides a scalable and performant solution, but retaining as much compatibility as possible with the Docker workflow. However, Shifter requires more administrative setup than the equivalent Charliecloud UDSS.

### F. Charliecloud

Charliecloud [7] developed at LANL to be a lightweight open source UDSS implementation based on the Linux user namespace for HPC sites with strict security requirements and consists of approximately 800 lines of code. It uses

Docker to build a UDSS image, shell scripts to unpack the image to an appropriate location and a C program to activate the image and run user code within the image.

In these secure environments, Charliecloud's distinct advantage is the usage of the newly introduced user namespace to support non-privileged launch of containerized applications. The user namespace is an unprivileged namespace and within the user namespace, all other privileged namespaces are created without the requirement of root privileges, which means that a containerized application can be launched without requiring privileged access to the host system.

Charliecloud employs Docker tools locally, but unlike Docker, the container is flattened to a single archive file in preparation for execution. To enable the distributed execution of tasks across multiple nodes, the containerized application is scaled by distributing the archive file to each compute node and unpacked into a local tmpfs environment.

*G. TensorFlow*

TensorFlow [8], developed and made available by Google, is a software framework to help developing and running Deep Learning based solutions.

TensorFlow is one of the most popular machine learning framework, and it has been used in a wide variety of applications and to conduct AI research. The developer can also deploy machine learning systems into production across numerous areas of computer science and other fields, including speech recognition, computer vision, robotics, and information retrieval, natural language processing and medical applications.

The framework is flexible and includes training and inference configurations for deep neural network models, which can be executed with minimal or no change on a wide variety of systems.

Tensorflow can be built from source either in a standard way or with external packages, that aims at improving performances, like the Intel® MKL library. However, this approach would require to pre-download all the packages in order to have them available during the build process. An additional load, that will be there regardless any plausible external package, is represented by Bazel: a building tool that is mandatory for producing Tensorflow Wheels. Ultimately, the release cadence of TensorFlow is significantly higher than most traditional HPC software packages and sometimes new versions are realized within the same months. This means that all modifications to the build process need to be redone every few weeks.

In order to take full advantage of Intel® architecture and to extract maximum performance, the TensorFlow framework has been optimized using Intel® Math Kernel Library for Deep Neural Networks (Intel® MKL-DNN) primitives, a popular performance library for deep learning applications. The installation process is described also in the following article [9].

*H. Horovod*

Horovod [10] developed at Uber for TensorFlow, Keras [11] and PyTorch [12]. It uses the Message Passing Interface (MPI) as mechanism for communication to allow multi-node training. It uses MPI concepts such as *Allgather* and *Allreduce* to handle the cross-replicas communication and weight updates. This is in contrast to the standard TensorFlow hierarchical architecture in which workers pass parameter updates to centralized parameter servers. Horovod is installed as a separate Python package, and to install and verify use the following commands 1-4:

$$\% \text{ pip install horovod} \tag{1}$$

$$\% \text{ python} \tag{2}$$

$$>>> \text{ import tensorflow as tf} \tag{3}$$

$$>>> \text{ import horovod.tensorflow as hvd} \tag{4}$$

Calling Horovod's API from the model script, a standard build of TensorFlow can be used to run distributed training with minimal source code change required in script to support distributed training with MPI.

III. SYSTEM AND COMPONENTS

*A. SuperMUC-NG*

The SuperMUC-NG (SNG) system at the Leibniz Supercomputing Center of the Bavarian Academy of Science (BADW-LRZ), as shown in figure 1, has a peak performance of 26.9 petaflops, which consists of 311,040 Intel® Xeon® Platinum 8174 (Skylake-SP) CPU cores and 719 terabytes of main memory. The Skylake cores are arranged into eight "thin" and one "fat" island. The "thin" islands consists of 792 nodes with each node contains 48 CPU cores and 96 Gigabytes of main memory. The "fat" island consists of 144 nodes with 48 CPU cores per node and 768 Gigabytes of main memory.

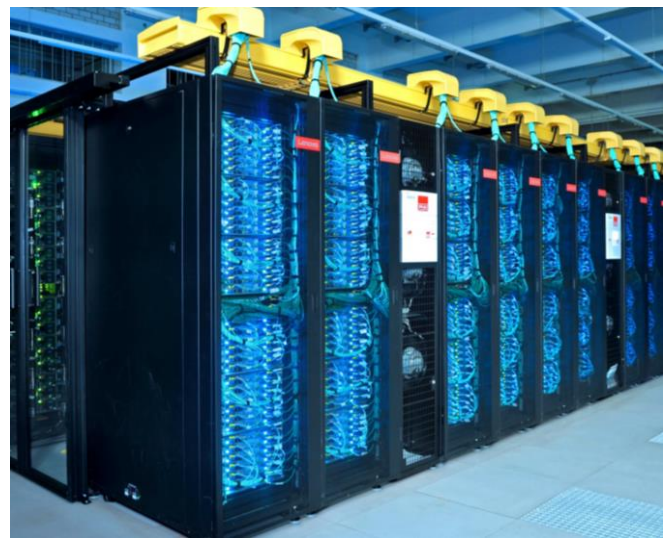

Fig. 1. SuperMUC-NG

All compute nodes within an individual island are connected via a fully non-blocking Intel Omnipath [13] OPA network with 100 Gbit/s. The Omnipath network topology is an inverted "fat" tree. The pruned interconnect enables a bidirectional bisection bandwidth ratio of 4:1 (intra-island / inter-island) for efficient communication. For high-performance I/O, the system employs, IBM's General Parallel File System (GPFS) [14] with 50 petabytes of capacity and an aggregated throughput of 500 gigabytes per second is available. In addition to a Data Science Storage (DSS) 20 petabytes with a throughput of 70 Gigabytes per second and a 256 terabyte "home" filesystem.

The operating system is SUSE Linux Enterprise Server (SLES) 12 SP3 [15]. Provisioning of the cluster is done with Extreme Cloud Administration Toolkit (xCAT) [16]. Job scheduling on the cluster is performed with the Slurm workload manager [17]. The systems base software stack is OpenHPC [18] compliant and the main development environment is Intel Parallel Studio XE 2019 & Intel MPI 2019.

SuperMUC-NG uses a novel form of "hot" water cooling developed at LRZ [19]. Active components like processors and memory are directly cooled with water that has an inlet temperature of ~40 degrees Celsius. The "hot" water cooling together with novel system software reduces the energy consumption of the system by up to 40%. The Lenovo SD650 DWC servers are designed to withstand an inlet water temperature of 50 degrees Celsius enabling 85-90 percent heat recovery. While the outlet water temperature is 60 degrees Celsius sent through an adsorption chiller, where it is converted to a chilled 20 degrees Celsius water suitable for cooling storage and networking components. Adsorption chilling will generate about 600 kilowatts of chilled water capacity. This translates into energy savings of more than 100,000 Euros per year.

For security reasons, SuperMUC-NG has no direct connection to the internet on both the login and compute nodes, and SSH has been disabled on the compute nodes.

### B. Charliecloud Conversion

To be able to deploy the modified Intel optimized Tensorflow Docker image on SuperMUC-NG it first must be converted into a Charliecloud container. Before we can proceed, Docker and Charliecloud must be installed on the developers Linux system, as some of the commands require privileged access.

First, we must build a container image from the dockerfile by executing the command ch-build as root in the directory of the dockerfile:

$$\text{desktop\% cd /path/to/dockerfile} \quad (5)$$

$$\text{desktop\% sudo ch-build --t <image\_name> <path>} \quad (6)$$

After a successful build, you will not see anything under your own directories, as ch-build hands control over to the Docker Daemon, which takes care of the image creation and places the image in the local Docker image registry. To verify that the image has been created successfully list the Docker images available with the command:

$$\text{desktop\% sudo docker images} \quad (7)$$

| REPOSITORY | TAG | IMAGE_ID | CREATED | SIZE |
|---|---|---|---|---|
| TensorFlow | HPC | a13c76f42e08 | 1 min ago | 4.05GB |

Now that you have a container image you can created a compressed tar archive of the image as a root. You can transfer the tar.gz archive file from your local system to the HPC system with the following commands.

$$\text{desktop\% sudo ch-docker2tar <image> <path\_tarfile>} \quad (8)$$

$$\text{desktop\% scp image.tar.gz user@cluster:<dest\_path>} \quad (9)$$

After successfully copying the compressed archive file to the HPC system the next step is to decompress and unpack the tar.gz file with the following command:

$$\text{cluster\% ch-tar2dir <path/image.tar.gz> <dest\_path>} \quad (10)$$

Make sure that you don't have a directory with the same name as your image, because the ch-tar2dir command will attempt to create and overwrite the existing directory. Now we can verify that the containerized image executes correctly by running the following commands 10-12:

$$\text{cluster\% module load charliecloud} \quad (10)$$

$$\text{cluster\% ch-run -w image -- echo "container hello world!"} \quad (11)$$

$$\text{container hello world!} \quad (12)$$

If you need to install software in the containerized image that requires access to another system, you must do this on a system that can access the required servers. For example, the command 'pip install' will not succeed because the system doesn't allow access via https.

The method we employed at LRZ was to download the Intel optimized TensorFlow Docker image from the Intel AI Docker Hub [20]. Then modified the containerized OS environment to support distributed training by installing MPI libraries with apt-get install and Horovod via pip install. This modified Docker image was saved and converted to a Charliecloud image using the command shown in step 8.

## IV. EXECUTION ON SUPERMUC-NG

In this section, we discuss how we deploy a Charliecloud container on SuperMUC-NG. In addition to describing the TensorFlow developed at CERN.

### A. The 3DGAN Network

Calorimeters are important components of HEP detectors. They are segmented in a large number of cells: by recording the energy deposited in each cell, it is possible to reconstruct the energy pattern produced by showers of secondary particles inside the detector volume. Because of their complexity, calorimeters are extremely demanding (in terms of time and computing resources) to simulate using classical Monte Carlo based approaches. However, by interpreting each cell as a pixel in an image and the amount of deposited energy per cell as the pixel grey-scale intensity, it is possible to represent the calorimeter output as a 3-dimensional image and apply Deep Learning models, developed typically for computer vision problems. 3DGAN leverages on this interpretation and uses 3D convolutional Generative Adversarial Networks to simulate synthetic energy showers, the same way they will be recorded by next-generation high granularity calorimeters [21]. In terms of physics, the initial validation of 3DGAN results shows a remarkable agreement with respect to state-of-the-art Monte Carlo-based simulation [22].

The training dataset is produced in the context of the current design studies of the Linear Collider Detector (LCD) for the Compact Linear Collider (CLIC), a concept for a future linear particle accelerator [23]. It consists of an electromagnetic calorimeters modelled as a regular grid of 3D cells of 5.1 mm3. Each entry in the dataset represents the energy depositions in individual calorimeter cells produced by one electron, stored as 25x25x25 pixel image [24]. Details about 3DGAN model can be found in [25]: it uses three-dimensional convolutions in order to capture the whole shower development along the three space dimensions. Loosely following an auxiliary classifier GAN approach [26] a generator and a discriminator network are trained according to an adversarial strategy [27], using physics quantities such as the particle type and its energy to condition the training and improve training stability. The 3DGAN model also implements a custom loss function; overall it sums up to slightly less than 1 million parameters [28]. The architecture is implemented using Keras and Tensorflow as a backend. The RMSProp [29] optimizer is used.

Integration with the Horovod framework and optimisation of Tensorflow and Horovod parameters in order to improve the training process performance on Intel Xeon processors is detailed in [29]: first results on scaling out the training process achieve close to 94% scaling efficiency up to 128 nodes.

*B. Slurm Single Node*

Deep learning networks are typically executed in an interactive manner on a workstation. However, interactive execute mode is not the standard way of executing applications on a HPC system, which employ a batch scheduler. One of the most popular job schedulers is Slurm, which is employed at LRZ.

Within a single shared memory node it is possible to employ both MPI and OpenMP to enable the parallel execution of tasks. However, for most cases OpenMP is considered to be the most appropriate for single node execution. The command to be added to the Slurm script to execute the application from within a Charliecloud container is:

ch-run my_container – name_of_executable <args>     (13)

*C. Slurm Multi-node*

On an HPC system the standard execution method involves distributing the computation across several nodes and MPI is the standard method used to enable the distribution of computation and parallel task execution across the nodes on the system.

An example of an execution command within Slurm submission script to enable the parallel execution of the application across multiple nodes from within the Charliecloud container is:

mpirun –n 4 ch-run my_container – mpi_hello_world     (14)

## V. RESULTS

In this section we discuss the results of deploying a Charliecloud container and executing parallel applications with MPI on SuperMUC-NG.

*A. Multi-node Scaling*

We were able to execute a distributed horovod enabled TensorFlow network on SuperMUC-NG. The training of the network was performed using hybrid MPI and OpenMP parallelism. That involved one MPI rank per node and two threads per physical CPU core to take advantage of hyperthreading. So that the network is executing on 96 hyperthreaded cores per node. All jobs were submitted to the system via the Slurm batch system.

TABLE I.

| Nodes | Time[S] per Epoch |
|---|---|
| 4 | 3806 |
| 8 | 1910 |
| 16 | 1001 |
| 32 | 504 |

Table 1 shows the time taken in seconds for a single epoch of the CERN network on 4, 8, 16 and 32 Intel Skylake nodes of SuperMUC-NG.

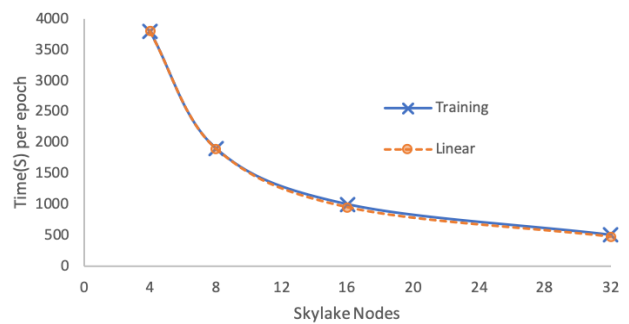

Fig. 2.  Scaling plot for the training of the TensorFlow on SuperMUC-NG

The scaling plot for the training of the CERN networks on SuperMUC-NG as illustrated by the blue line compared to the linear scaling orange line in figure 2, shows that the training exhibits almost linear scaling up to 32 Intel Skylake nodes.

*B. Overheads*

Today HPC systems come with limited amount of resource (i.e. compute, system memory, storage). It is important that newly introduced software stacks, such as the applied Charliecloud approach, exhibit minimal overhead in terms of runtime and utilization of the system resources. For example, the amount of system memory on a single compute node in HPC can be limited and exceeding it, which can cause the job to crash. As shown in Table 2., we observe no performance overhead for AI workloads when using Charliecloud for AI in running topologies such as ResNet-50 and AlexNet.

TABLE II.

| Benchmark | TF Throughput with Charliecloud [img/s] | TF Throughput without Charliecloud [img/s] |
|---|---|---|
| AlexNet with cifar10 | 1968 | 1973 |
| ResNet-50 | 75 | 74 |

Table 2. Achieved throughput [img/s] for AlexNet and ResNet-50 based on Tensorflow 1.11 with and without Charliecloud.

We measured small neglible memory overhead when running the workloads through the Charliecloud runtime environment (Table 3).

TABLE III.

| Benchmark | Free System Memory with Charliecloud [GB] | Free System Memory without Charliecloud [GB] |
|---|---|---|
| AlexNet with cifar | 331.29 | 331.33 |
| ResNet50 with imagenet | 324.47 | 324.89 |

Table 3. Free system memory for AlexNet and Resnet50 based on Tensorflow 1.11 with and without Charliecloud.

## VI. CONCLUSIONS

We were able to set-up the software stack, implement and test the full experimental pipeline in less than one day by using Intel optimized version of TF and Intel mpi. The results are extremely promising and justify the effort we have made in developing a secure containerization solution for deploying AI on HPC systems.

## FUTURE WORK

This work is to be intended as proof of concept for deploying AI on a secure HPC system, and we plan to build a BKM (Best-Known-Method) to share for production-ready deployments.

On the technical side, we would also like to extend this work to investigate the deployment other AI frameworks, and validate procedure on new workloads.

## ACKNOWLEDGMENT

The authors would like to thank The Leibniz Supercomputing Centre Bavarian Academy of Sciences and Humanities for granting us access to SuperMUC-NG.